\documentclass[aps,prb,reprint,amssymb,amsmath,superscriptaddress,floatfix]{revtex4-1}

\usepackage{graphicx}
\usepackage{epstopdf}
\usepackage[english]{babel}
\bibpunct{[}{]}{,}{n}{}{}

\begin{document}

\title{Negative  polarizability  of 2D electrons in  HgTe quantum well}

\author{V.\,Ya.~Aleshkin}
\affiliation{Institute for Physics of Microstructures  RAS, 603087 Nizhny Novgorod, Russia}

\affiliation{Lobachevsky University of Nizhny Novgorod, 603950
Nizhny Novgorod, Russia}

\author{A.\,V.~Germanenko}

\affiliation{School of Natural Sciences and Mathematics, Ural Federal University,
620002 Ekaterinburg, Russia}

\author{G.\,M.~Minkov}
\affiliation{School of Natural Sciences and Mathematics, Ural Federal University,
620002 Ekaterinburg, Russia}

\affiliation{M.N. Miheev Institute of Metal Physicsx of Ural Branch of Russian Academy of Sciences, 620108 Ekaterinburg, Russia}

\author{A.\,A.~Sherstobitov}

\affiliation{School of Natural Sciences and Mathematics, Ural Federal University,
620002 Ekaterinburg, Russia}

\affiliation{M.N. Miheev Institute of Metal Physicsx of Ural Branch of Russian Academy of Sciences, 620108 Ekaterinburg, Russia}

\date{\today}

\begin{abstract}
The polarizability of electrons occupying the lowest subband of spatial quantization in CdTe/Cd$_x$Hg$_{1-x}$Te/CdTe quantum wells is calculated. It is shown that polarizability in the quantum well  without cadmium is negative, i.e., the displacement of an electron in an electric field applied perpendicularly to the quantum well plane is opposite to the force acting on it. The negative polarizability of 2D electrons can reduce the dielectric constant of quantum wells by up to $(10-15)$ percent.
\end{abstract}


\maketitle

\section{Introduction}
\label{sec:intr}
Two-dimensional (2D) systems based on gapless semiconductors HgTe are unique
object. Mercury telluride is semiconductor with inverted ordering of $\Gamma_6$ and $\Gamma_8$ bands. The $\Gamma_6$ band, which is the conduction band in conventional semiconductor, is located in HgTe lower in the energy than the $\Gamma_8$ band. Unusual positioning of the bands leads to crucial features of the electron spectrum under space
confinement \cite{Volkov76,Dyakonov82,Lin85,Gerchikov90}. So
at some critical width of the HgTe/CdTe quantum well (QW), $d= d_c\simeq 6.3$~nm, the energy spectrum is gapless and linear \cite{Bernevig06}.  In the wide quantum wells, $d>d_c$, the lowest electron subband is mainly formed from the $\Gamma_8$ states at small quasimomentum value ($k$), while the $\Gamma_6$ states form the hole states in the depth of the valence band. Such a band structure is referred to as inverted
structure. At $d<d_c$, the band ordering is normal; the highest valence
subband at zero quasimomentum is formed from the heavy hole $\Gamma_8$ states, while the lowest electron subband is formed both from the $\Gamma_6$ states and from the light  $\Gamma_8$ states.

These peculiarities manifest themselves in a new and interesting physical phenomena as exemplified by topological states arising in such  systems (see, review paper \cite{Ren2016}). In the present paper we report one further anomalous effect originated from the specific of the electron spectrum in the HgTe QWs.

It is well known that electrons occupying the lowest subband of spatial quantization in  a quantum well of semiconductors with a normal band structure possess positive polarizability with respect to an electric field applying along the normal to the quantum well. For instance,  an electron in a GaAs quantum well  shifts in the direction of force under influence of an electric field. Recall the polarizability ($\alpha$) determines the response of a bound system to external field. It is the coupling coefficient between the dipole moment $\mathbf{P}$ arising in  the electric field $\mathbf{E}$;  $\mathbf{P}=\alpha \mathbf{E}$. Polarizability is often used to describe the behavior of molecules and atoms in an electric field \cite{Bonin}. But it can also be used to describe electrons in quantum wells \cite{Ando82}. As will be shown below, electrons in the lowest subband in HgTe/CdTe quantum well subjected to an electric field are displaced in the direction opposite to the acting force. This means that 2D electron gas in the HgTe/CdTe wells has a negative polarizability. This effect is predicted in the HgTe QWs of different width both with inverted and with normal energy spectrum. Therewith, the polarizability increases in the Cd$_x$Hg$_{1-x}$Te/CdTe quantum wells with the increasing cadmium content and  becomes positive at $x>0.168$ when the band ordering of the bulk Hg$_{1-x}$Cd$_{c}$Te becomes normal

\begin{figure}
\includegraphics[width=0.99\linewidth,clip=true]{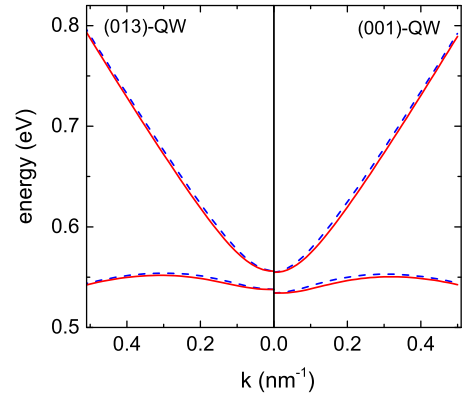}
\caption{The  $E$~vs~$k$ dependences for the $(013)$- and  $(001)$-HgTe/CdTe QWs of $12$~nm width. The $k$ vector is directed along  $[100]$. The solid and dashed curves are two branches split due to the presence of electric field and interface inversion asymmetry. }
\label{f1}
\end{figure}

\section{Results and discussion}
To calculate the polarizability, the ensemble average displacement of the mean electron coordinate was calculated. The calculations were carried out in the framework of the Kane model taking into account the deformation effects and the absence of inversion symmetry of the interfaces forming the quantum well. As  known an additional term to the Hamiltonian describing symmetry lowering on heterointerfaces was suggested by E.L.Ivchenko (see Ref.~\cite{Born}). This term is given in the Appendix. The parameters of materials and the method of solving the Schr\"{o}dinger equation  were taken from \cite{Minkov17}. As an example, Fig.~\ref{f1} shows the electron spectrum in  HgTe/CdTe QW of $12$~nm width grown on the $(001)$ and $(013)$ planes in the presence of a uniform electric field $E=10$~kV/cm applied along the normal to the quantum well, which coincides with \emph{z}-axis. The interface inversion asymmetry  and the electric field lead to a small spin splitting of the subbands. Note the conduction band spectrum  is almost isotropic and close to each other for both substrate orientations.

\begin{figure}
\includegraphics[width=0.9\linewidth,clip=true]{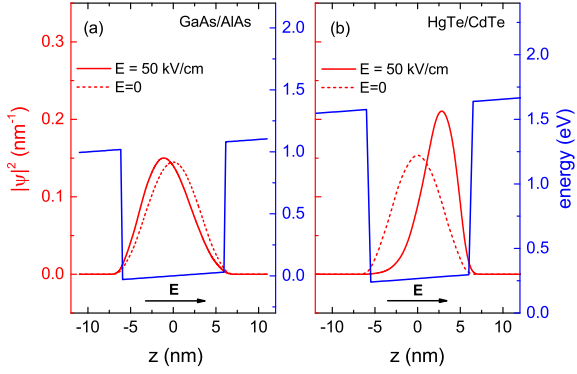}
\caption{(Color online) The dependence $|\psi(z)|^2$ in the presence  and absence of electric field for the GaAs/AlAs (a)  and HgTe/CdTe (b) quantum wells, $k=0$. The run of the conduction band bottom in the presence of electric field are shown as well.}
\label{f2}
\end{figure}

In order to demonstrate an unusual response of electrons in the lowest subband in HgTe/CdTe QW to the electric field, let us compare this effect in GaAs and HgTe based quantum wells. Figure~\ref{f2} shows the probability density $|\psi(z)|^2$ for the electron on the bottom of conduction band in GaAs/AlAs and HgTe/CdTe QWs  in the presence and absence of the electric field. As clearly seen the mean electron coordinate shifts in opposite directions in GaAs/AlAs and HgTe/CdTe quantum wells  when the electric field is applied.
In the GaAs/AlAs quantum well the electron density shifts against the electric field [Fig.~\ref{f2}(a)], i.e., in the direction of the force $-e\mathbf{E}$ ($e$ is the elementary charge) which exerts the electron. In the HgTe/CdTe quantum well, the electron shifts against the force [Fig.~\ref{f2}(b)] that is in contradiction with intuitive expectations.

\begin{figure}
\includegraphics[width=\linewidth,clip=true]{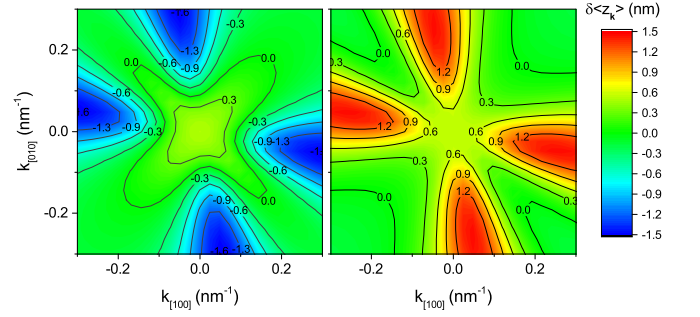}
\caption{(Color online) The  quasimomentum dependence of the electron shift
$\delta \langle z_\mathbf{k}^-\rangle$ and $\delta \langle z_\mathbf{k}^+\rangle$ for the lower (left panel) and upper (right panel) spin-split branches, respectively,   in the  (001)-HgTe/CdTe quantum well subjected to electric field $10$~kV/cm.}
\label{f3}
\end{figure}

\begin{figure}
\includegraphics[width=\linewidth,clip=true]{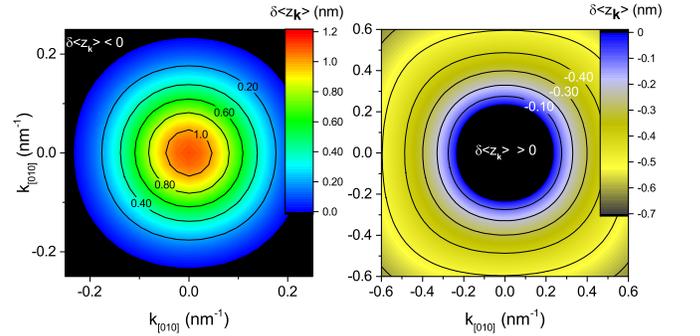}
\caption{(Color online) The  quasimomentum dependence of the net shift $\delta \langle z_\mathbf{k}\rangle= \delta \langle z_\mathbf{k}^-\rangle+\delta \langle z_\mathbf{k}^+\rangle$ obtained by summing the data presented in the left and right panels of Fig.~\ref{f3}. The left and right panels show the data for $\delta \langle z_\mathbf{k}\rangle >0$ and $\delta \langle z_\mathbf{k}\rangle<0$, respectively. }
\label{f31}
\end{figure}

Our calculations show that the shift of electron density in the HgTe QWs is strongly dependent on the quasimomentum vector.  This is illustrated by Fig.~\ref{f3}. It shows the $\mathbf{k}$ dependences of the change in the electron mean coordinate  calculated as
\begin{equation}
\delta \langle z_\mathbf{k}^\pm\rangle=\langle z_\mathbf{k}^\pm(E)\rangle-\langle z_\mathbf{k}^\pm(0)\rangle,
\label{eq01}
\end{equation}
where
\begin{equation}
\langle z_\mathbf{k}^\pm\rangle =\int_{-\infty}^\infty z |\psi_\mathbf{k}^\pm(z)|^2 dz,
\label{eq02}
\end{equation}
in the electric field of $10$~kV/cm for the lowest conduction subband of spatial quantization in (100)-HgTe/CdTe QW. The superscripts  $+$ and $-$  correspond to two split branches. Positive $\delta\langle z_\mathbf{k}^\pm\rangle$ values  correspond to the shift of the electron in the direction which is opposite to the force exerting on the electron by the electric field. As seen from Fig.~\ref{f3} the $\delta\langle z_\mathbf{k}^\pm\rangle$ value depends  on the direction and absolute value of the wave vector drastically. But, what is more important, electrons belonging to  different spin branches have $\delta \langle z_\mathbf{k}^+\rangle$ and $\delta \langle z_\mathbf{k}^-\rangle$ of different  sign, i.e., they are shifted in the opposite directions upon the electric field.  Note the quatrefoils in Fig.~\ref{f3} are not of forth order symmetry as would be expected in the [001] quantum well. This is due to the presence of interface inversion asymmetry which is taken into account in our consideration as mentioned above. At the same time the $\mathbf{k}$ dependence of the total for two spin states shift $\delta \langle z_\mathbf{k}\rangle= \delta \langle z_\mathbf{k}^-\rangle + \delta \langle z_\mathbf{k}^+\rangle$ exhibits nearly  cylindrical symmetry as seen from Fig.~\ref{f31}. From the same figure is also evident that $\delta \langle z_\mathbf{k}\rangle$  behaves itself anomaly only at low $k$ values $k \lesssim 0.23$~nm$^{-1}$ (the left panel in Fig.~\ref{f31}), while for $k \gtrsim 0.23$~nm$^{-1}$ it demonstrate normal behavior characterized by $\delta \langle z_\mathbf{k}\rangle<0$  (see the right panel in Fig.~\ref{f31}) like that in conventional QWs. Analogous picture is emerged for the (013)-HgTe/CdTe QWs.

Let  us now consider how  the electron polarizability affects the dielectric constant ($\varepsilon$) of the quantum well. In a bulk material, the dielectric constant can be defined as the ratio of the electric induction to the electric field. The electric induction is sum of the electric field and polarization of a unit volume multiplied by 4$\pi$\cite{Tamm}. In a quantum well additional polarization occurs due to electron presence in the conduction band. Average over quantum well value of this polarization is $\alpha En/d$, where $E$ is the electric field value, $n$ is the sheet electron concentration, $d$ is the quantum well width, $\alpha$ is the electron polarizability. Therefore we obtain the following expression for the average dielectric constant

\begin{equation}
\varepsilon = \varepsilon_i + \frac{4\pi\alpha n}{d},
\label{eq04}
\end{equation}
where  $\varepsilon_i$  is the dielectric constant of the quantum well without electrons  (equal to $20.0$ and $14.4$ for the low and high frequencies, respectively \cite{Rogalski05}) and
\begin{equation}
\alpha=-e\delta \langle\langle z \rangle\rangle/E.
\label{eq05}
\end{equation}
Here, $\delta \langle\langle z \rangle\rangle$ is change in the electron mean coordinate, Eq.~(\ref{eq02}), averaged over the ensemble of 2D electrons.

\begin{figure}
\includegraphics[width=0.99\linewidth,clip=true]{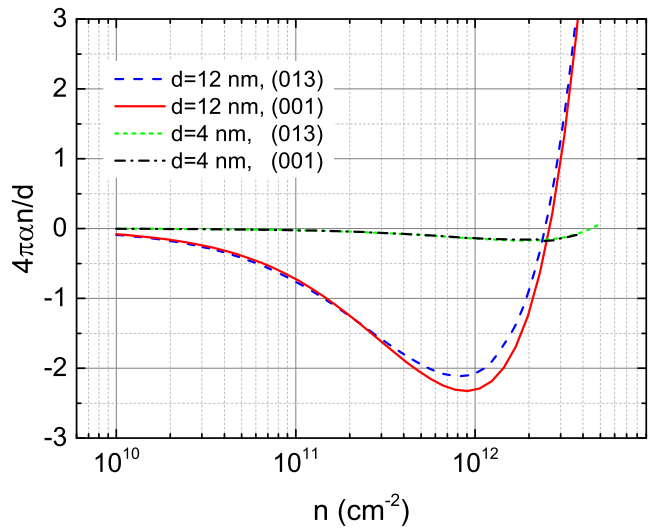}
\caption{(Color online) The electron density dependence of the electron polarizability  contribution to the dielectric constant of the quantum wells with $d=4.0$~nm and $12$~nm grown on (001) and (013) planes.}
\label{f4}
\end{figure}

Fig.~\ref{f4} shows the electron density dependences of the polarizability contribution $4\pi\alpha n/d$ for (001)- and (013)-HgTe/CdTe QWs of two different widths, $d=12$~nm and $4$~nm, which correspond to the inverted and normal energy spectrum, respectively.  As seen the electrons have a negative polarizability in the quantum wells both with the  inverted and with the  normal band structure. It is also seen that the absolute value of the electron contribution to the dielectric constant of $12$~nm QW exceeds that for the  $4$~nm quantum well. Note, the calculations gives approximately the same results when the barriers are the solid solution Cd$_{0.6}$Hg$_{0.4}$Te, i.e., the change of the height of the barriers within reasonable limits does not affect the polarizability.

In order to clarify the physical cause of negative electron polarizability, we note three important arguments. The first argument is that the wave functions near $k = 0$ in 12~nm and 4~nm quantum wells are formed mainly by the states of the $\Gamma_8$ band.  The proportion of these states in the electron wave function decreases with increasing wave vector, similarly to a decrease in the polarization magnitude (see Fig.~\ref{f31}).
The second argument is that the electron polarizability is positive in such quantum wells in which the electron wave function is formed mainly by the states of the $\Gamma_6$ band. For examples, $1$~nm-HgTe/CdTe quantum well or quantum wells with a cadmium fraction greater than $0.168$ can be given. The third argument is that the $\Gamma_8$-band states form the electron wave functions of the valence band in the GaAs/AlAs quantum wells mentioned above. The electron polarizability of the valence band states in such quantum wells is {\it negative}, which corresponds to a {\it positive} hole polarizability. Taking into account these arguments, we can conclude that the cause of the negative electron polarizability in HgTe/CdTe QWs is the presence of a significant fraction of the $\Gamma_8$ band states in the electron wave function.

If the QW heterostructure is doped asymmetrically,  an intrinsic electric  field  exists in the well even  in the absence of an external field due to the spatial separation of electrons and ionized donors. In order to estimate  the effect in this case, we have calculated the polarizability of the electron gas in the quantum well subjected to uniform field of $50$~kV/cm. It turns out that the $n$ dependence of the electron polarizability  contribution to the dielectric constant remains  qualitatively the same, but its absolute value is approximately $10$ percent less as compared with  the rectangular quantum well.

\section{Conclusion}

Electrons in the conduction band of  the HgTe quantum well exhibit an amazing behavior in an electric field normal to the quantum well plane. The electron density is shifted in the  direction opposite to the acting force. As a result, the electron polarizability of  2D electron gas is negative over a wide range of electron densities. The negative polarizability of a 2D electrons  reduces the dielectric constant of the rectangular HgTe quantum well on the maximal value of about $(10 - 15)$ percent at $n\simeq 10^{12}$~cm$^{-2}$. The effect weakens with  increasing Cd content in the quantum well and the polarizability becomes positive at $x\simeq 0.168$ when the bulk Cd$_x$Hg$_{1-x}$Te spectrum  transforms from the gapless with inverted band structure to the spectrum with open gap and normal band ordering. The anomalous negative polarizability can be detected through the Kerr and/or Pockels effects, it can manifests itself in the capacitance experiments performed on the gated heterostructures with HgTe/CdTe quantum wells.

\begin{acknowledgments}

The work has been supported in part by the Russian Foundation for Basic
Research (Grant \#18-02-00050), by  Act 211 Government of the Russian Federation, agreement \#02.A03.21.0006,  by  the Ministry of Education and Science of the Russian Federation under Projects \#0035-2019-0020-C-01 and \#3.9534.2017/8.9.

\end{acknowledgments}

\appendix*
\section{The Ivchenko term}

We use the Ivchenko term responsible for the interface inversion asymmetry  in the following form
\begin{eqnarray}
H_I&=&\pm \gamma_4\delta(z\pm d/2) \nonumber \\
&\times &\left(
\begin{array}{cccccccc}
0& 0& 0& 0& 0& 0& 0& 0\\
0& 0& 0& 0& 0& 0& 0& 0\\
0& 0& 0& 0& -i& 0& 0& i\sqrt{2}\\
0& 0& 0& 0& 0& -i& 0& 0\\
0& 0& i&0& 0& 0& 0& 0\\
0& 0& 0& i& 0& 0& i\sqrt{2}& 0\\
0& 0& 0& 0& 0& -i\sqrt{2}& 0& 0\\
0& 0& -i\sqrt{2}& 0& 0& 0& 0& 0
\end{array}
\right)
\end{eqnarray}
The value of $\gamma_4$ was chosen as $8$~eV/nm  in order to agree with results from Ref.~\cite{Tarasenko15}.

%

\end{document}